\documentclass[showpacs,
               showkeys,preprintnumbers,amsmath,amssymb]{revtex4}

\usepackage{graphicx}
\usepackage{dcolumn}
\usepackage{bm}
\def\nns{$g_{N\!N\sigma}$\,}
\def\bea{\begin{eqnarray}}
\def\eea{\end{eqnarray}}
\def\nnb{\nonumber}
\def\qqbar{\langle \bar{q}q\rangle}
\def\lamt{\tilde{\lambda}_N}

\begin{document}


\title{The nucleon-sigma coupling constant in QCD Sum Rules}

\author{G. Erkol}
\email{erkol@kvi.nl}
\author{R. G. E. Timmermans}%
\email{timmermans@kvi.nl}
\affiliation{
          Theory Group, KVI \\
          University of Groningen \\
          Groningen, The Netherlands}
\author{Th. A. Rijken}
\email{t.rijken@science.ru.nl}
\affiliation{
          Institute for Theoretical Physics \\
          Radboud University Nijmegen \\
          Nijmegen, The Netherlands}

\date{\today}

\begin{abstract}
The external-field QCD Sum Rules method is used to evaluate the coupling
constant of the light isoscalar-scalar meson (``$\sigma$'' or $\varepsilon$)
to the nucleon. The contributions that come from the excited nucleon states
and the response of the continuum threshold to the external field are
calculated. The obtained value of the coupling constant is compatible with
the large value required in one-boson exchange potential models of the
two-nucleon interaction.
\end{abstract}

\maketitle

\section{Introduction}
The values of the meson-baryon coupling constants are of particular
interest in understanding the nucleon-nucleon (NN)~\cite{Nag78,Sto94}
and hyperon-nucleon (YN)~\cite{Mae89,Rij98} interactions
in terms of {\em e.g.} one-boson exchange (OBE) models. The scalar mesons
play a significant role in such phenomenological potential models. The
structure and even the status of the scalar mesons have, however, always
been controversial~\cite{Tim94,Swa94}.
In early OBE models for the NN interaction the exchange of an isoscalar-scalar
``$\sigma$'' meson with a mass of about 500 MeV was needed to obtain enough
medium-range attraction and a sufficiently strong spin-orbit force. It was
only later understood that the exchange of a broad isoscalar-scalar meson,
the $\varepsilon$(760), simulates the exchange of such a low-mass
``$\sigma$''~\cite{Bin72}.
The $\varepsilon$(760) is difficult to detect because it is broad and hidden
under the strong signal from the $\rho^0$(770). There are strong arguments
from chiral symmetry for the existence of such a light isoscalar-scalar meson
approximately degenerate with the $\rho$ meson~\cite{Wei90}.

In the quark model, the simplest assumption for the structure of the scalar
mesons is the $^3P_0$ $q\bar{q}$ states. In this case, the scalar mesons
might form a complete nonet of dressed $q\bar{q}$ states, resulting from
{\em e.g.} the coupling of the $P$-wave $q\bar{q}$ states to meson-meson
channels~\cite{Bev86}. Explicitly, the unitary singlet and octet states,
denoted respectively by $\varepsilon_1$ and $\varepsilon_8$, read
\bea\label{octsing}
   \varepsilon_1 & = & (u\bar{u}+d\bar{d}+s\bar{s})/\sqrt{3} \, , \nnb \\
   \varepsilon_8 & = & (u\bar{u}+d\bar{d}-2s\bar{s})/\sqrt{6}\, .
\eea 
The physical states are mixtures of the pure $SU(3)$-flavor states,
and are written as
\bea\label{mix}
   \varepsilon & = & \cos\theta_s\,\varepsilon_1+\sin\theta_s\,\varepsilon_8
                     \, , \nnb \\
           f_0 & = &-\sin\theta_s\,\varepsilon_1+\cos\theta_s\,\varepsilon_8 
                     \, .
\eea
For ideal mixing holds that $\tan\theta_s=1/\sqrt{2}$ or
$\theta_s\simeq 35.3^\circ$, and thus one would identify
\bea\label{psqqb}
   \varepsilon(760) & = & (u\bar{u}+d\bar{d})/\sqrt{2}\, , \nnb \\
           f_0(980) & = & -s\bar{s} \, .
\eea
The isotriplet member of the octet is $a_0^{\pm,0}$(980), where
\bea\label{azero}
   a_0^0(980) = (u\bar{u}-d\bar{d})/\sqrt{2} \, .
\eea

An alternative and arguably
more natural explanation for the masses and decay properties of the
lightest scalar mesons is to regard these as cryptoexotic $q^2\bar{q}^2$
states~\cite{Jaf77}. In the MIT bag model, the scalar $q\bar{q}$ states are
predicted around $1250$ MeV, while the attractive color-magnetic force results
in a low-lying nonet of scalar $q^2\bar{q}^2$ mesons~\cite{Jaf77,Aer80}.
This nonet contains a nearly degenerate set of $I=0$ and $I=1$ states, which
are identified as the $f_0(980)$ and $a^{\pm,0}_0(980)$ at the $\bar{K}K$
threshold, where
\bea\label{ps4q}
   a^0_0(980) & = & (sd\bar{s}\bar{d}-su\bar{s}\bar{u})/\sqrt{2} \, , \nnb \\
     f_0(980) & = & (sd\bar{s}\bar{d}+su\bar{s}\bar{u})/\sqrt{2} \, ,
\eea
with the ideal-mixing angle $\tan\theta_s=-\sqrt{2}$ or $\theta_s\simeq
 -54.8^\circ$ in this case. The light isoscalar member of the nonet is
\bea\label{sigmaq4}
   \varepsilon(760) & = & ud\bar{u}\bar{d} \, .
\eea
The nonet is completed by the strange member $\kappa$(880), which like the
$\varepsilon$(760) is difficult to detect because it is hidden under the
strong signal from the $K^*$(892)~\cite{Tim94,Swa94}. In keeping with
other recent works~\cite{Bla99,Mai04,Bri04} we will use in this paper the
nomenclature $(a_0^{\pm,0},f_0,\sigma,\kappa)$ for the scalar-meson nonet,
where one should identify $\sigma=\varepsilon(760)$, but we will not rely
on a particular theoretical prejudice about the quark structure of the
light scalar mesons.

One way to make progress with the scalar mesons is to study their role
in the various two-baryon reactions (NN, YN, YY). Our aim in this paper
is to calculate the nucleon-$\sigma$ coupling constant \nns\ by using the
QCD Sum Rules (QCDSR) method~\cite{Shi79}. QCDSR links the hadronic degrees
of freedom with the underlying QCD parameters, and serves as a powerful
tool to extract qualitative and quantitative information about hadron
properties~\cite{Rei84,Col00}. In this framework, one starts with a
correlation function that is constructed in terms of hadron interpolating
fields. On the theoretical side, the correlation function is calculated
using the Operator Product Expansion (OPE) in the Euclidian region. This
correlation function is matched with an {\em Ansatz} that is introduced in
terms of hadronic degrees of freedom on the phenomenological side. The
matching provides a determination of hadronic parameters like baryon
masses, magnetic moments, coupling constants of hadrons, and so on.

The QCDSR method has been extensively used to investigate
meson-baryon coupling constants. One usually starts with the
vacuum-to-vacuum matrix element of the correlation function that
is constructed with the interpolating fields of two baryons and
one meson. However, this three-point function method has as a major
drawback that at low momentum transfer the OPE fails.
Moreover, when the momentum of the meson is large, the latter is
plagued by problems with higher resonance contamination~\cite{Mal97}.
A method that can be used at low momentum transfer is the external-field
method~\cite{Iof84}. There are two formulations that can be used to
construct the external-field sum rules:
The first one is to start with a vacuum-to-vacuum transition matrix element
of the nucleon interpolating fields. In this approach, no vacuum-to-meson
matrix elements occur, but one has to know the response of the various
condensates in the vacuum to the external field, which can be described by
a susceptibility $\chi$. This method has been used to determine the magnetic
moments of baryons~\cite{Iof84,Bal83,Chi86,Chi85}, the nucleon axial
coupling constant~\cite{Chi85,Bel84}, the nucleon sigma term~\cite{Jin93},
and baryon isospin mass splittings~\cite{Jin95}. In the second approach,
one starts with a vacuum-to-meson transition matrix element of the nucleon
interpolating fields, where some other transition matrix elements
should be evaluated~\cite{Rei84}. (This is also the
starting point of the light-cone QCDSR method.) In~\cite{Shi95}, 
pion-nucleon coupling constant was calculated in the soft meson limit using 
this approach. Later it was pointed out that the sum rule for pion-nucleon coupling
in the soft-meson limit can be reduced to the sum rule for the nucleon mass 
by a chiral rotation so the coupling was calculated again with a finite meson
momentum~\cite{Bir96}. These calculations were improved considering the coupling schemes 
at different Dirac structures and beyond the chiral limit 
contributions~\cite{Oka98,Lee98,Kim99}. This coupling constant has also 
been calculated using the vacuum-to-vacuum method~\cite{Hwa97,Hwa96}, 
and it was pointed out that the sum rule that is constructed 
for the coupling is independent and it is not reduced to the nucleon mass 
sum rule by a chiral rotation.

In this paper, we calculate the nucleon-$\sigma$ coupling constant \nns\
by using the external-field QCDSR method. We evaluate the vacuum-to-vacuum
transition matrix element of the two-nucleon interpolating fields
in an external isoscalar-scalar field, and construct two sum rules, one
of which leads to a stable result with respect to variations in the Borel
mass. We also compute the contributions that come from the excited
nucleon states and the response of the continuum threshold to the
external field. Previously, the strong and weak (parity-violating)
pion-nucleon coupling constants~\cite{Hwa97,Hen96} and the coupling
constants of the vector mesons $\rho$(770) and $\omega$(782) to the 
nucleon~\cite{Wen97} were calculated by using this method.

We will compare our result for the coupling constant with the value from
a successful OBE model of the NN interaction, the Nijmegen soft-core
potential~\cite{Nag78,Sto94}, which was originally derived from Regge-pole
theory. (The coupling constants of this OBE model
were analyzed from the point of view of the large-$N_c$ expansion of
QCD in Ref.~\cite{Kap97}.) It is then important to realize that in NN
potential models the coupling constants of the heavy mesons to the nucleon
are determined by the (``non-peripheral'') $S$-, $P$-, and $D$-waves.
Therefore, the fits to the scattering data are sensitive to {\em e.g.}
the volume integral of the OBE potentials, which
is proportional to the coupling at $t=0$~\cite{Swa78} ($t=-p^2$, where
$p$ is the four-momentum of the meson). The coupling constants obtained from
the external-field QCDSR method are also defined at $t=0$, and therefore the
comparison to the OBE model is appropriate.

Our paper is organized as follows: In Section~\ref{secNSR} we present the
formulation of QCDSR with an external isoscalar-scalar field and construct
the relevant sum rules. We give the numerical analysis of the sum rules and
discuss the results in Section~\ref{secAN}. Finally, we arrive at our
conclusions in Section~\ref{secCONC}.

\section{Nucleon Sum Rules in an external scalar field}\label{secNSR}
In the external-field QCDSR method one starts with the correlation
function of the nucleon interpolating fields in the presence of an
external constant isoscalar-scalar field $\sigma$, defined by
\bea\label{cor1}
   \Pi^\sigma(q) = i\int d^4 x~ e^{i q\cdot x}\, \Big \langle 0\Big
         |{\cal T}[\eta_N(x)\bar{\eta}_N(0)]\Big |0\Big\rangle_\sigma \,\, ,
\eea
where $\eta_N$ is the Ioffe nucleon
interpolating field~\cite{Iof81}
\bea\label{intfi}
   \eta_N = \epsilon_{abc}[u_a^T C\gamma_\mu u_b]\gamma_5\gamma^\mu d_c \,\, .
\eea
Here $a,b,c$ denote the color indices, and $T$
and $C$ denote transposition and charge conjugation, respectively.
The external scalar field contributes to the correlation function
in Eq. (\ref{cor1}) in two ways: First, it directly couples the
quark field in the nucleon current and second, it modifies the
condensates by polarizing the QCD vacuum. In the presence of an
external scalar field there are no correlators that break
Lorentz invariance, like $\langle\bar{q}\sigma_{\mu\nu}q\rangle$
which appears in the case of an external
electromagnetic field $F^{\mu\nu}$. However, the correlators already
existing in the vacuum are modified by the external field, {\em viz}.
\bea\label{vaccon}
   \qqbar_\sigma &\equiv& \qqbar
          + g^\sigma_q \chi \sigma \qqbar \, , \nnb \\ \langle g_c \bar{q}
          {\bm\sigma}\cdot {\bm G} q\rangle_\sigma &\equiv& \langle g_c
          \bar{q} {\bm\sigma}\cdot {\bm G} q\rangle + g^\sigma_q \chi_G
          \sigma \langle g_c \bar{q} {\bm\sigma}\cdot {\bm G} q\rangle \, ,
\eea
where $g^q_\sigma$ is the quark-$\sigma$ coupling constant and, $\chi$ and $\chi_G$ are the susceptibilities corresponding to quark and mixed quark-gluon condensates,
respectively. We have assumed that the responses of the up
and the down quarks to the external (isoscalar) field are the same.

In the Euclidian region, the OPE of the product of two
interpolating fields can be written as
\bea\label{opex}
      \Pi^\sigma(q)=\sum_{n} C^\sigma_n(q) O_n \,\, ,
\eea
where $C^\sigma_n(q)$ are the Wilson coefficients and $O_n$ are the local
operators in terms of quark and gluon fields. At the quark level, we have
\bea\label{cor2}
   \Big \langle 0\Big |{\cal
   T}[\eta_N(x)\bar{\eta}_N(0)]\Big |0\Big \rangle_\sigma= 2 i
   \epsilon^{abc}\epsilon^{a^\prime b^\prime c^\prime} {\rm Tr} \{S_u^{b
   b^\prime}(x) \gamma_\nu C [S_u^{a a^\prime}(x)]^T C
   \gamma_\mu\}\gamma_5 \gamma^\mu S_d^{c
   c^\prime}(x)\gamma^\nu\gamma_5 \, .
\eea
In order to calculate the
Wilson coefficients, we need the quark propagator in the presence
of the external sigma field. In coordinate space the full quark
propagator takes the form
\bea\label{proptot}
   S_q(x)=S_q^{(0)}(x)+S_q^{(\sigma)}(x) \, ,
\eea
where
\bea\label{prop0}
    i~S_q^{(0)ab}&\equiv&\langle 0|T[q^a(x)
    \bar{q}^b(0)|0\rangle_0 \nnb \\ &=&
    \frac{i~\delta^{ab}}{2\pi^2x^4}\hat{x}-\frac{i~\lambda_{ab}^n}{32
    \pi^2}\frac{g_c}{2} G_{\mu\nu}^n
    \frac{1}{x^2}(\sigma^{\mu\nu}\hat{x}+\hat{x}
    \sigma^{\mu\nu})-\frac{\delta^{ab}}{12}\langle\bar{q}q\rangle-\frac{\delta^{ab}x^2}{192}\langle
    g_c \bar{q}{\bm\sigma}\cdot {\bm G} q\rangle \, ,
\eea
and
\bea\label{propE}
   i~S_q^{(\sigma)ab} &\equiv&\langle 0|T[q^a(x)
  \bar{q}^b(0)|0\rangle_\sigma \nnb \\ &=&g^\sigma_q \sigma
  \Big[-\frac{ \delta^{ab}}{4\pi^2
  x^2}-\frac{1}{32\pi^2}\lambda_{ab}^n g_c
  G_{\mu\nu}^n\sigma^{\mu\nu} \ln(-x^2)-\frac{\delta^{ab}\langle
  g_c^2 G^2\rangle}{2^9\times 3 \pi^2} x^2
  \ln(-x^2) \nnb\\ &&~~~~~~+\frac{i~\delta^{ab}}{48}\qqbar\hat{x}-\frac{\delta^{ab}\chi}{12}\qqbar+
  \frac{i~\delta^{ab}x^2}{2^7\times 3^2}\langle g_c
  \bar{q}{\bm\sigma} \cdot {\bm G}
  q\rangle\hat{x} \nnb\\ &&~~~~~~ -\frac{\delta^{ab}x^2}{192}
  \chi_G \langle g_c \bar{q}{\bm\sigma}\cdot {\bm G} q\rangle\Big ] \, .
\eea
In these expressions, $G^{\mu\nu}$ is the gluon field tensor and 
$g_c^2=4\pi\alpha_s$ is the quark-gluon coupling constant squared. We do not include
terms that are proportional to the quark masses, since these terms give
negligible contributions to the final result.

\begin{figure}[h]
\includegraphics[scale=0.70]{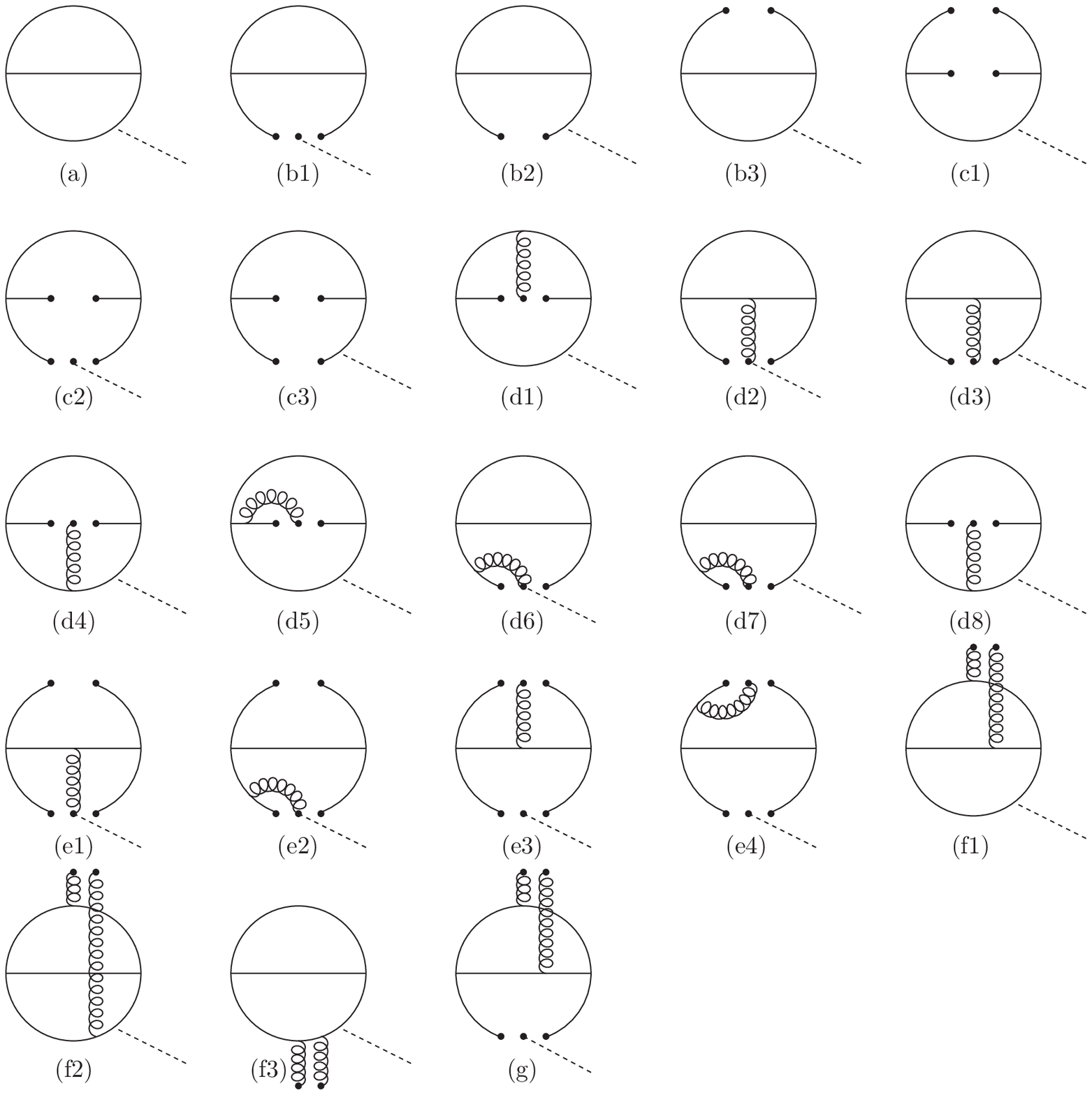}
\caption{The diagrams that were used to calculate the Wilson coefficients
         of the correlation functions $\Pi_\sigma^q$ and $\Pi_\sigma^1$. The solid, wavy, and
         the dashed lines represent the quark, gluon, and the external
         scalar field, respectively.}
\label{diag}\end{figure}

Using the quark propagator in Eq.~(\ref{proptot}), one can compute
the correlation function $\Pi^\sigma(q)$. The diagrams that we use to
calculate the Wilson coefficients of $\Pi^\sigma(q)$ are shown in
Fig.~\ref{diag}. Lorentz covariance
and parity imply that $\Pi^\sigma(q)$ takes the form
\bea
  \Pi^\sigma(q) = ( \Pi_\sigma^1 + \Pi_\sigma^q \hat{q} ) \sigma + ( \Pi_0^1 + \Pi_0^q \hat{q} ) \, ,
\eea
where $\hat{q} = q^\mu \gamma_\mu$. Here $\Pi_0^1$ and $\Pi_0^q$ represent the invariant functions in the vicinity of the external field, which can be used to construct the mass sum rules for the nucleon, and $\Pi_\sigma^1$ and $\Pi_\sigma^q$ denote the invariant functions in the presence of the external field. Using these invariant functions, one can derive the sum rules at the structures $1$ and $\hat{q}$. $\Pi_\sigma^q$ and $\Pi_\sigma^1$ are evaluated as follows:
\bea\label{qmom}
   \Pi^q_\sigma(q)= g_q^\sigma
   \frac{1}{(2\pi)^4}\Big [a_q\,
   \ln(-q^2)-\chi\,\frac{4}{3q^2}\,a_q^2+\frac{m_0^2}{2q^2}a_q
      -(\chi+\chi_G)\,\frac{m_0^2}{6q^4}\,a_q^2 \Big ] \, ,
\eea
and
\bea\label{1mom}
   \Pi^1_\sigma(q)= g_q^\sigma
   \frac{1}{(2\pi)^4}&&\Big [-\frac{q^4}{2}
   \ln(-q^2)-\frac{10}{3q^2}a_q^2-\chi\, a_q\, q^2 \ln(-q^2)+\chi_G
   \frac{m_0^2}{2} a_q\, \ln(-q^2) \nnb \\
   &&~~+\frac{b}{8} \ln(-q^2)-\chi
   \frac{b}{24 q^2} a_q \Big ] \, ,
\eea
where we have defined
$a_q\equiv-(2\pi)^2\qqbar$, $b\equiv\langle g_c^2 G^2 \rangle$, and
$m_0^2\equiv\langle g_c\bar{q}{\bm\sigma}\cdot{\bm G}q\rangle/ \qqbar$.

We now turn to the calculation of the hadronic side.
We saturate the correlator in Eq.~(\ref{cor1})
with nucleon states and write
\bea\label{sat}
   \Pi^\sigma(q) = \frac{\langle 0|\eta_N|N\rangle}{q^2-M_N^2}\,\,
   \langle N|\sigma N\rangle \,\,\frac{\langle N| \bar{\eta}_N | 0
   \rangle}{q^2-M_N^2} \,\, ,
\eea
where $M_N$ is the mass of the
nucleon. The matrix element of the current $\eta_N$ between the
vacuum and the nucleon state is defined as
\bea\label{overlap}
   \langle 0 | \eta_N | N \rangle= \lambda_N \upsilon \,\,,
\eea
where $\lambda_N$ is the overlap amplitude and $\upsilon$ is the Dirac
spinor for the nucleon, normalized as
$\bar{\upsilon}\upsilon=2M_N$. Inserting Eq.~(\ref{overlap}) into
Eq.~(\ref{sat}) and defining \nns\ via the interaction Lagrangian density
\bea
   {\cal L} = -g_{N\!N\sigma}\,
              \bar{\upsilon} \upsilon\sigma \,\, ,
\eea
we obtain for the hadronic part
\bea\label{phpart}
   -|\lambda_N|^2\frac{\hat{q}+M_N}{q^2-M_N^2} g_{N\!N\sigma}
   \frac{\hat{q}+M_N}{q^2-M_N^2} \,\, .
\eea

In addition, there are contributions coming from the excitations to higher
nucleon states which are written as
\bea\label{phpartex}
   -\lambda_N\lambda_{N^\ast}\,\frac{\hat{q}+M_N}{q^2-M_N^2}\, g_{N\!N^\ast
   \sigma}\,\frac{\hat{q}+M_{N^\ast}}{q^2-M_{N^\ast}^2} \,\, ,
\eea
as well as contributions coming from the intermediate states due to
$\sigma$-$N$ scattering, {\em i.e.} the {\em continuum} contributions. The
term that corresponds to the excitations to higher nucleon states
also has a pole at the nucleon mass, but a single pole instead of
a double one like in Eq.~(\ref{phpart}). This single-pole term is
not ``damped'' after the Borel transformation and should be included
in the calculations.

Finally, there is another contribution that comes from the response
of the continuum to the external field, given by
\bea\label{cont}
   \int^\infty_0 \frac{-\Delta s_0 ~b(s)}{s-q^2}
   \delta(s-s_0) ds \, ,
\eea
where $s_0$ is the continuum threshold,
$\Delta s_0$ is the response of the continuum threshold to the
external field, and $b(s)$ is a function that is calculated from the
OPE. When $\Delta s_0$ is large, this term should also be included
in the hadronic part~\cite{Iof95}.

The QCD sum rules are obtained by matching the OPE side with the
hadronic side and applying the Borel transformation. The resulting
sum rules are:
\bea\label{sumq}
   &\Big[&-M^4\, a_q\, E_0+\frac{4}{3}\, \chi\, M^2\,
   a_q^2\,L^{4/9}-\frac{m_0^2}{2}\, M^2\,
   a_q\,L^{-14/27}-(\chi+\chi_G)\,\frac{m_0^2}{6}\,a_q^2\, L^{-2/27}
   \Big]~e^{M_N^2/M^2} \nnb \\ &&
   ~~~~~~=-\lamt^2\, \frac{M_N}{g_q^\sigma}\, g_{N\!N\sigma}
   + \tilde{B}_q\, \frac{M^2}{g_q^\sigma}+\frac{(s^q_0)^2}{2g_q^\sigma}
   \,\Delta s^q_0\, M^2\,L^{-4/9} e^{(M_N^2-s_0^q)/M^2}\, ,
\eea
and
\bea\label{sum1}
   &\Big[& 2\,M^8\,E_2\,L^{-4/9}+\frac{20}{3}\,M^2\,a_q^2\,L^{4/9}+2\, \chi\,
   a_q\, M^6\, E_1-\chi_G\, m_0^2\, a_q\, M^4\, E_0\, L^{-14/27} \nnb \\
   &&-\frac{b}{4}\, M^4\, E_0\,L^{-4/9}+\chi\,
   \frac{b}{12}\, M^2\, a_q 
   \Big]~e^{M_N^2/M^2} \nnb \\ &&~~~~~~=-(2
   M_N^2-M^2)\,\frac{\lamt^2}{g_q^\sigma}\, g_{N\!N\sigma}
   + \tilde{B}_1\, \frac{M^2}{g_q^\sigma} + \frac{4}{g_q^\sigma}\, 
   a_q\, s^1_0\, \Delta
   s^1_0\, M^2\, e^{(M_N^2-s_0^1)/M^2}\, ,
\eea
where $M$ is the Borel mass and we have
defined $\lamt^2=32 \pi^4 \lambda_N^2$.
The continuum contributions are included by the factors
\bea
  E_0&\equiv& 1- e^{-s_0^i/M^2}\, ,\nnb\\E_1&\equiv& 1-
  e^{-s_0^i/M^2}\Big(1+\frac{s_0^i}{M^2}\Big)\, ,\nnb\\E_2&\equiv&
  1-e^{-s_0^i/M^2}\Big(1+\frac{s_0^i}{M^2}+\frac{(s_0^i)^2}{2M^4}\Big)\, ,
\eea
where $s_0^i$ are the continuum thresholds with $i=1,q$. In
the sum rules above, we have included the single-pole
contributions with the factors $\tilde{B}_i$. The third terms on the
right-hand side (RHS) of Eqs.~(\ref{sumq}) and (\ref{sum1}) denote the
contributions that are explained in Eq.~(\ref{cont}).
These terms are suppressed by the factor $e^{-(s_0^i-M_N^2)/M^2}$
as compared to the single-pole terms. We have incorporated the effects
of the anomalous dimensions of the various operators through the factor
$L=\ln(M^2/\Lambda_{QCD}^2)/\ln(\mu^2/\Lambda_{QCD}^2)$.

\section{Analysis of the sum rules and discussion}\label{secAN}
In this Section we analyze the sum rules derived in the previous
Section in order to determine the value of \nns. We observe that
the sum rule in Eq.~(\ref{sumq}) is more stable than the other sum
rule in Eq.~(\ref{sum1}), so we use only this sum rule for the
numerical analysis. Such a comparison and conclusion have been
made about these sum rules also in some earlier
works~\cite{Jin93,Jin95}.

In order to calculate \nns, we need to know the values of the
scalar susceptibilities $\chi$ and $\chi_G$. The value of the
susceptibility $\chi$ can be calculated by using the two-point
function~\cite{Jin93}
\bea\label{susc}
   T(p^2)=i\int d^4 x
   e^{i p\cdot x} \Big\langle 0\Big|{\cal T}[\bar{u}(x)u(x)+\bar{d}(x)
   d(x),\bar{u}(0)u(0)+\bar{d}(0) d(0)]\Big|0\Big\rangle\, ,
\eea
via the relation
\bea\label{suscorr}
   \chi \qqbar=\frac{1}{2} T(0) .
\eea
The two-point function in Eq.~(\ref{susc}) at $p^2=0$ has been calculated
in chiral perturbation theory~\cite{Gas83} with the result
\bea
   \chi=\frac{\qqbar}{16\pi^2 f_\pi^4}\Big(\frac{2}{3}
   \bar{\ell}_1+\frac{7}{3}\bar{\ell}_2-\frac{11}{6}\Big)\, ,
\eea
where $f_\pi=93$ MeV is the pion decay constant and $\bar{\ell}_1$
and $\bar{\ell}_2$ are low-energy constants appearing in the
effective chiral Lagrangian. The values of these low-energy
constants have been estimated previously in various works (see {\em e.g.}
Ref.~\cite{Col01} for a review). A recent analysis of
$\pi$-$\pi$ scattering gives $\bar{\ell}_1= -1.9\pm 0.2$ and
$\bar{\ell}_2=5.25\pm 0.04$~\cite{Col01}, which is
consistent with earlier determinations, but with smaller
uncertainties. Using these values of $\bar{\ell}_1$ and
$\bar{\ell}_2$ and taking the quark condensate $a_q=0.51\pm 0.03$
GeV$^3$, we find $\chi=-10\pm 1$ GeV$^{-1}$. The value of
the susceptibility $\chi_G$ is less certain. Therefore, we
allow $\chi_G$ to vary in a wider range. We also adopt
$b=4.7\times 10^{-1}$ GeV$^4$, $\lamt^2=2.1$ GeV$^6$, and
$m_0^2=0.8$ GeV$^2$~\cite{Iof84,Ovc88}. We take
$M_N=0.94$ GeV, the renormalization scale $\mu=0.5$ GeV,
and the QCD scale parameter $\Lambda_{QCD}=0.1$ GeV.
It is relevant to point out that the choice of the two-point
function in Eq.~(\ref{susc}) does not imply a theoretical prejudice
about the structure of the scalar mesons. What is calculated is
just the susceptibility pertaining to the response of the quark
condensates $\qqbar$ to the scalar $q\bar{q}$ field, as shown in
Eq.~(\ref{vaccon}).

To proceed to the numerical analysis, we arrange the RHS of Eq.~(\ref{sumq})
in the form
\bea\label{form}
    f(M^2) = A_q + B_q M^2+C_q M^2 L^{-4/9} e^{(M_N^2-s_0^q)/M^2} \ ,
\eea
and fit the left-hand side (LHS) to $f(M^2)$. Here we have defined
\bea\label{form2}
   A_q &\equiv& -\lamt^2 \frac{M_N}{g_q^\sigma} g_{N\!N\sigma} \, , \nnb \\
   B_q &\equiv& \frac{\tilde{B}_q}{g_q^\sigma} \, , \nnb \\
   C_q &\equiv& \frac{(s^q_0)^2}{2g_q^\sigma} \Delta s^q_0 \, .
\eea

\begin{figure}[h]
\includegraphics[scale=0.85]{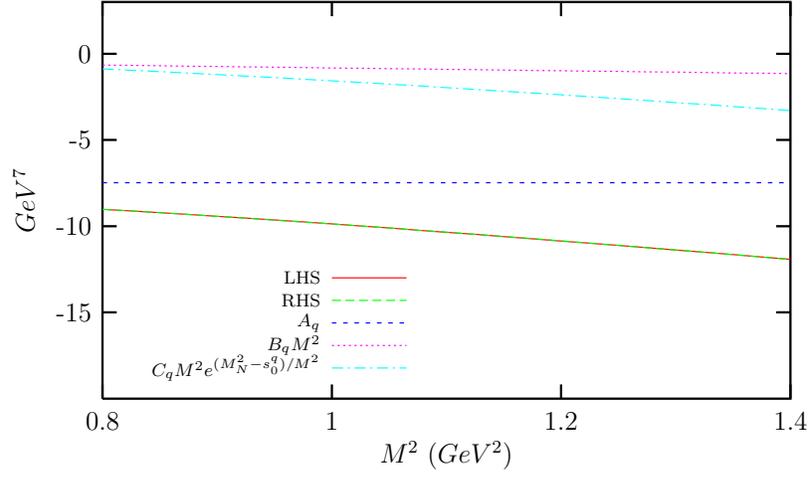}
\caption{(Color online) The Borel mass dependence of LHS and the fitted RHS of
         Eq.~(\ref{sumq}) for $s^q_0=2.3$ GeV$^2$ and $\chi_G\equiv\chi=-10$
         GeV$^{-1}$. We also present the terms on the RHS individually.
         Note that the LHS
         curve (solid) overlies the RHS curve (dashed).}
\label{fig1}\end{figure}

In Fig.~\ref{fig1}, we present the Borel mass dependence of the
LHS and the RHS of Eq.~(\ref{sumq}) for $s^q_0=2.3$ and $\chi_G
\equiv\chi=-10$ GeV$^{-1}$. We choose the Borel window 0.8 GeV$^2$
$\leq M^2\leq 1.4$ GeV$^2$ which is commonly identified as the fiducial
region for the nucleon mass sum rules~\cite{Iof84}. It is seen that
the LHS curve (solid) overlies the RHS curve (dashed). In order to
estimate the contributions that come from the excited nucleon
states and the response of the continuum threshold, we plot each
term on the RHS individually. We observe that the single-pole
terms (dotted) give very small contribution, but the response of
the continuum threshold (dot-dashed) is quite sizable. Nevertheless,
the summation of these curves with the line of the double-pole term
(small-dashed) gives a stable sum rule.

\begin{figure}[h]
\includegraphics[scale=0.85]{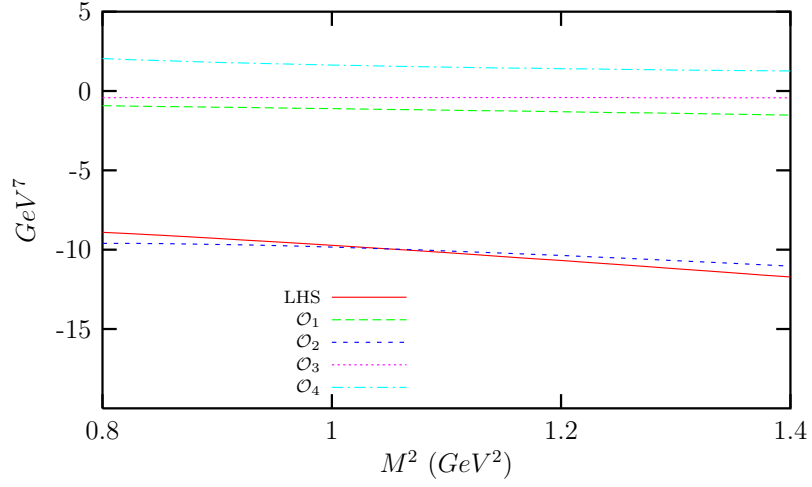}
\caption{(Color online) The four terms on the LHS of Eq.~(\ref{sumq}) individually,
         together with the summation of them for $s^q_0=2.3$ GeV$^2$ and
         $\chi_G\equiv\chi=-10$ GeV$^{-1}$. Here ${\cal O}_1$ denotes the
         first term, ${\cal O}_2$ denotes the second term, and so on.}
\label{fig2}\end{figure}

In Fig.~\ref{fig2}, we plot the Borel mass dependence of the four terms
on the LHS of Eq.~(\ref{sumq}) separately, together with their summation
for $s^q_0=2.3$ GeV$^2$ and $\chi_G
\equiv\chi=-10$ GeV$^{-1}$. This helps us
to compare the contributions of different operators on the OPE side.
Here ${\cal O}_1$ denotes the first term, ${\cal O}_2$ denotes the second
term, and so on. We observe that ${\cal O}_1$ and ${\cal O}_3$ are small,
${\cal O}_4$ is sizable, and ${\cal O}_2$ is large. The term ${\cal O}_4$
contributes with different sign with respect to ${\cal O}_1$ and ${\cal O}_3$,
and so tends to cancel the latter. Therefore \nns\ is mainly determined by
${\cal O}_2$ on the LHS.

\begin{figure}[h]
\includegraphics[scale=0.85]{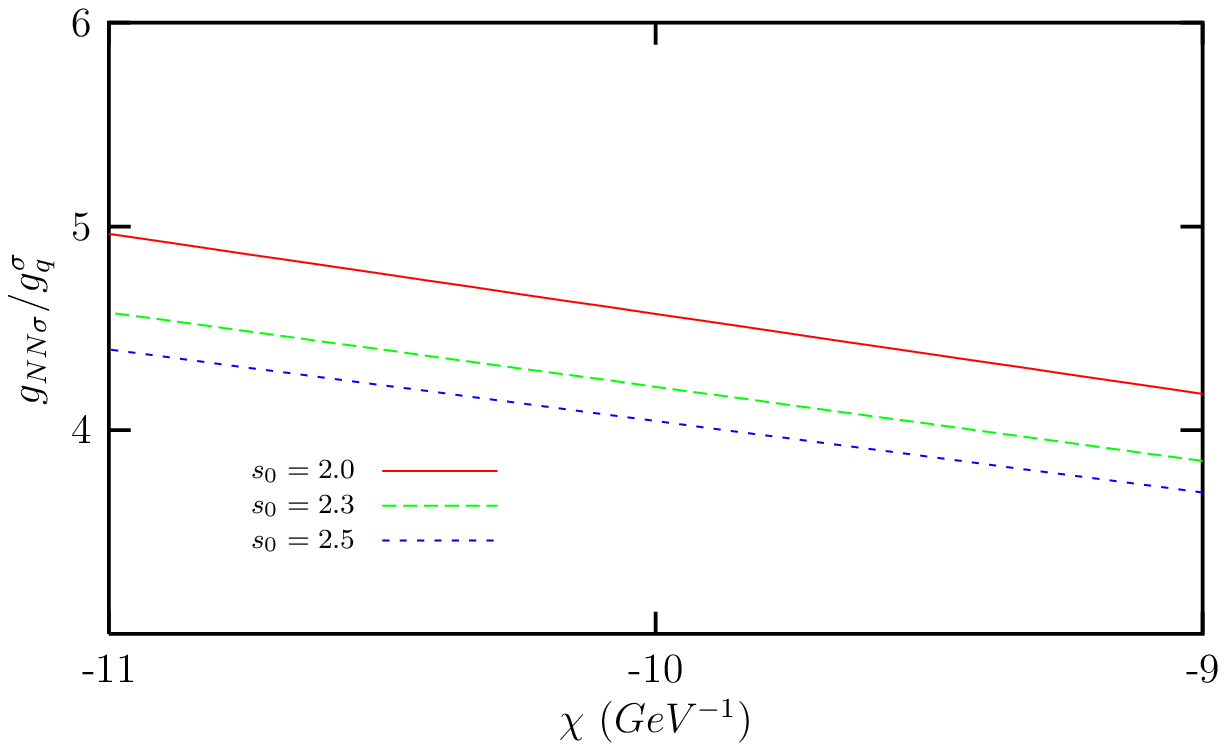}
\caption{(Color online) The dependence of $g_{N\!N\sigma}/g_q^\sigma$ on the susceptibility
         $\chi$ for three different values of $s_0^q=2.0$, 2.3, and 2.5
        GeV$^2$; here we take $\chi\equiv\chi_G$.}
\label{fig3}\end{figure}

In order to see the sensitivity of the coupling constant on the continuum
threshold and the susceptibility $\chi$, we plot in Fig.~\ref{fig3}
the dependence of $g_{N\!N\sigma}/g_q^\sigma$ on $\chi$ for three different
values $s_0^q=2.0$, 2.3, and 2.5 GeV$^2$, and taking $\chi\equiv\chi_G$.
One sees that \nns\ changes by
approximately $8\%$ in the considered region of the
susceptibility $\chi$. The value of \nns\ is not very sensitive to
a change in $s_0^q$, which gives an uncertainty of approximately
$6\%$ to the final value. Taking into account the uncertainty in $\chi$,
$s_0^q$, and $a_q$, the predicted value for $g_{N\!N\sigma}/g_q^\sigma$
of the sum rule in Eq.~(\ref{sumq}) reads
\bea\label{nns}
   g_{N\!N\sigma}/g_q^\sigma = 3.9 \pm 1.0 \, .
\eea

In a similar way, one can calculate the other two terms on the
RHS of Eq. (\ref{sumq}) as: \bea\label{other}B_q&=&-0.2\pm 1.2
~\text{GeV}^5\nnb\, ,\\C_q&=&-7.9\pm 2.9 ~\text{GeV}^5\,.\eea As
noted above, the value of the susceptibility $\chi_G$ is less
certain than the value of $\chi$. If we let $\chi_G$ change in a
wider range, say $6 ~\text{GeV}^{-1} \leq -\chi_G \leq 14
~\text{GeV}^{-1}$, this brings an additional $15 \%$ uncertainty
to the value quoted in Eq. (\ref{nns}).

The ratio in Eq. (\ref{nns}) is in agreement with the naive quark model,
which gives $g_{N\!N\sigma}/g_q^\sigma=3$ based on counting the $u$- and
the $d$-quarks in the nucleon. (Ideal mixing in the scalar sector is assumed
above, that is, the sigma meson is taken without a strange-quark content.)
Another estimate can be made from the ratio of pion-nucleon to pion-quark
coupling constant, $g_{N\!N\pi}/g_q^\pi$. Since the $\sigma$ meson is the
chiral partner of the pion~\cite{Wei90}, one expects 
\bea\label{rat}
   g_{N\!N\sigma}/g_q^\sigma = g_{N\!N\pi}/g_q^\pi \, .
\eea 
Using the Goldberger-Treiman relation for both the pion-nucleon
and the pion-constituent quark couplings,
\bea\label{GT}
   g_{N\!N\pi} & = & g^A_N\frac{M_N}{f_\pi} \, , \nnb \\
   g_q^\pi   & = & g^A_q\frac{m_q}{f_\pi} \, ,
\eea
where $m_q$ is the mass of the constituent quark, $g^A_N$ and $g^A_q$
are the nucleon and the quark axial couplings, respectively, one
obtains~\cite{Glo95}
\bea\label{GTr}
   \frac{g_{N\!N\pi}}{g_q^\pi} = \frac{5}{3}\frac{M_N}{m_q} \, .
\eea
With a constituent-quark mass of 340 MeV~\cite{Glo95},
Eq.~(\ref{GTr}) yields $g_{N\!N\pi}/g_q^\pi=4.6$. Using Eq.~(\ref{rat})
we find that this agrees nicely with the QCDSR result in Eq.~(\ref{nns}). 

To determine \nns, one next has to assume some value for the quark-$\sigma$
coupling constant $g_q^\sigma$. Adopting the value $g_q^\sigma=3.7$ as
estimated from the sigma model~\cite{Ris99}, we obtain
\bea\label{sigNNf}
   g_{N\!N\sigma} = 14.4 \pm 3.7 \, .
\eea
The coupling constant in Eq.~(\ref{sigNNf}) is defined at $t=0$, {\em i.e.}
$g_{N\!N\sigma} \equiv g_{N\!N\sigma}(t=0)$. As stressed above, also in NN
potential models the heavy-meson coupling constants are determined at $t=0$.
The (large) value of \nns\ obtained in Eq.~(\ref{sigNNf}) is in agreement
with the value $g_{N\!N\sigma}=16.9$ from the Nijmegen soft-core NN potential
model~\cite{Nag78}, obtained from a fit to the NN scattering data.

\section{Conclusion}\label{secCONC}
We have calculated the coupling constant \nns\ of the isoscalar-scalar
meson, which plays a significant role in OBE models of the NN and YN
interactions, to the nucleon, using the external-field QCDSR method.
Our main result is the ratio $g_{N\!N\sigma}/g_q^\sigma$ in Eq.~(\ref{nns})
which is determined purely from QCDSR. The value of \nns\ is dependent
on $g_q^\sigma$, the value of which we use as estimated in the sigma model.
The obtained value of \nns\ is in agreement with the large value found
in OBE models. We have also computed the contributions that
come from the excited nucleon states and the response of the
continuum threshold to the external field. We observe that while
the single-pole contributions are small, the response of the
continuum threshold is sizable. We plan to extend the
external-field QCDSR method to the hyperons and the complete scalar-meson
nonet, in order to address the $SU(3)$-flavor structure of the scalar-meson
coupling constants to the baryon octet~\cite{Erkol}.

\begin{acknowledgments}
We are grateful to M. Nielsen and M. Oka for useful discussions and comments.
\end{acknowledgments}

\newpage

\begin{thebibliography}{100}
\bibitem{Nag78} M.~M.~Nagels, T.~A.~Rijken, and J.~J.~de Swart,
                Phys. Rev. D {\bf 17}, 768 (1978).
\bibitem{Sto94} V. G. J. Stoks, R. A. M. Klomp, C. P. F. Terheggen,
                and J. J. de Swart, Phys. Rev. C {\bf 49}, 2950 (1994).
\bibitem{Mae89} P.~M.~M.~Maessen, Th.~A.~Rijken, and J.~J.~de Swart,
                Phys. Rev. C {\bf 40}, 2226 (1989).
\bibitem{Rij98} Th.~A.~Rijken, V.~G.~J.~Stoks, and Y.~Yamamoto,
                Phys. Rev. C {\bf 59}, 21 (1999) [arXiv:nucl-th/9807082].
\bibitem{Tim94} R. G. E.~Timmermans, Th.~A.~Rijken, and J.~J.~de Swart,
                Phys. Rev. C {\bf 50}, 48 (1994) [arXiv:nucl-th/9403011].
\bibitem{Swa94} J. J. de Swart, P. M. M. Maessen, and T. A. Rijken,
                in: {\em Properties and Interactions of Hyperons},
                edited by B. F. Gibson, P. D. Barnes, and K. Nakai
                (World Scientific, Singapore, 1994), pp. 37-54
                [arXiv:nucl-th/9405008].
\bibitem{Bin72} J. Binstock and R. Bryan, Phys. Rev. D {\bf 4}, 1341 (1971);
                R. A. Bryan and A. Gersten, Phys. Rev. D {\bf 6}, 341 (1972).
\bibitem{Wei90} S. Weinberg, Phys. Rev. Lett. {\bf 65}, 1177 (1990).

\bibitem{Bev86} E.~van Beveren, T.~A.~Rijken, K.~Metzger, C.~Dullemond,
                G. Rupp, and J.~E.~Ribeiro,
                Z. Phys. C {\bf 30}, 615 (1986);
                E. van Beveren, G. Rupp, T.~A.~Rijken, and C.~Dullemond,
                Phys. Rev. D {\bf 27}, 1527 (1983).
\bibitem{Jaf77} R.~L.~Jaffe,
                Phys. Rev. D {\bf 15}, 281 (1977);
                Phys. Rev. D {\bf 17}, 1444 (1978).
\bibitem{Aer80} A.~T.~M.~Aerts, P.~J.~Mulders, and J.~J.~de Swart,
                Phys. Rev. D {\bf 21}, 1370 (1980).
\bibitem{Bla99} D. Black, A. H. Faziborz, F. Sannino, and J. Schechter,
                Phys. Rev. D {\bf 59}, 074026 (1999) [arXiv:hep-ph/9808415].
\bibitem{Mai04} L. Maiani, F. Piccinini, A. D. Polosa, and V. Riquer,
                Phys. Rev. Lett. {\bf 93}, 212002 (2004) [arXiv:hep-ph/0407017].
\bibitem{Bri04} T.~V.~Brito, F.~S.~Navarra, M.~Nielsen and M.~E.~Bracco,
  		Phys. Lett. B {\bf 608}, 69 (2005)
  		[arXiv:hep-ph/0411233].
\bibitem{Shi79} M.~A.~Shifman, A.~I.~Vainshtein, and V.~I.~Zakharov,
                Nucl. Phys. {\bf B147}, 385 (1979);
                Nucl. Phys. {\bf B147}, 448 (1979).
\bibitem{Rei84} L.~J.~Reinders, H.~Rubinstein, and S.~Yazaki,
                Phys. Rept. {\bf 127}, 1 (1985).
\bibitem{Col00} P.~Colangelo and A.~Khodjamirian, in: \emph{Boris Ioffe Festschrift 'At the Frontier of 			Particle Physics / Handbook of QCD' vol.3}, edited by M. Shifman (World Scientific, Singapore, 		2001).
\bibitem{Mal97} K. Maltman,
                Phys. Rev. C {\bf 57}, 69 (1998) [arXiv:hep-ph/9707231].
\bibitem{Iof84} B.~L.~Ioffe and A.~V.~Smilga,
                Nucl. Phys. {\bf B232}, 109 (1984).
\bibitem{Bal83} I.~I.~Balitsky and A.~V.~Yung,
                Phys. Lett. B {\bf 129}, 328 (1983).
\bibitem{Chi86} C.~B.~Chiu, J.~Pasupathy, and S.~L.~Wilson,
                Phys. Rev. D {\bf 33}, 1961 (1986).
\bibitem{Chi85} C.~B.~Chiu, J.~Pasupathy, and S.~L.~Wilson,
                Phys. Rev. D {\bf 32}, 1786 (1985).
\bibitem{Bel84} V.~M.~Belyaev and Y.~I.~Kogan,
                Phys. Lett. B {\bf 136}, 273 (1984).
\bibitem{Jin93} X.~M.~Jin, M.~Nielsen, and J.~Pasupathy,
                Phys. Lett. B {\bf 314}, 163 (1993).
\bibitem{Jin95} X.~M.~Jin,
                Phys. Rev. D {\bf 52}, 2964 (1995) [arXiv:hep-ph/9506299];
                X.~M.~Jin, M.~Nielsen, and J.~Pasupathy,
                Phys. Rev. D {\bf 51}, 3688 (1995) [arXiv:hep-ph/9405202].
\bibitem{Shi95} H.~Shiomi and T.~Hatsuda,
                Nucl. Phys. {\bf A594}, 294 (1995)
                [arXiv:hep-ph/9504354].
\bibitem{Bir96} M.~C.~Birse and B.~Krippa,
                Phys. Rev. C {\bf 54}, 3240 (1996) [arXiv:hep-ph/9606471].	
\bibitem{Oka98} H.~C.~Kim, S.~H.~Lee and M.~Oka,
  		Phys.\ Lett.\ B {\bf 453} (1999) 199 [arXiv:nucl-th/9809004].
\bibitem{Lee98} H.~C.~Kim, S.~H.~Lee and M.~Oka,
  		Phys.\ Rev.\ D {\bf 60} 034007 (1999) [arXiv:nucl-th/9811096].
\bibitem{Kim99} H.~C.~Kim, T.~Doi, M.~Oka, and S.~H.~Lee,
                Nucl. Phys. {\bf A662}, 371 (2000) [arXiv:nucl-th/9909007];
                Nucl. Phys. {\bf A678}, 295 (2000) [arXiv:nucl-th/0002011].
\bibitem{Hwa97} W.~Y.~Hwang,
                Z. Phys. C {\bf 75}, 701 (1997) [arXiv:hep-ph/9601219].
\bibitem{Hwa96} W.~Y.~Hwang, Z.~s.~Yang, Y.~S.~Zhong, Z.~N.~Zhou and S.~L.~Zhu,
  		Phys. Rev. C {\bf 57}, 61 (1998) [arXiv:nucl-th/9610025].
\bibitem{Hen96} E.~M.~Henley, W.~Y.~P.~Hwang, and L.~S.~Kisslinger,
                Phys. Lett. B {\bf 367}, 21 (1996);
                {\em ibid.}  B {\bf 440}, 449(E) (1998)
                [arXiv:nucl-th/9511002].
\bibitem{Wen97} Y.~Wen and W.~Y.~P.~Hwang,
                Phys. Rev. C {\bf 56}, 3346 (1997).

\bibitem{Kap97} D.~B.~Kaplan and A.~V.~Manohar,
                Phys. Rev. C {\bf 56}, 76 (1997)
                [arXiv:nucl-th/9612021].
\bibitem{Swa78} J.~J.~de Swart and M.~M.~Nagels,
                Fortsch. Phys. {\bf 26}, 215 (1978).

\bibitem{Iof81} B.~L.~Ioffe,
                Nucl. Phys. {\bf B188}, 317 (1981); {\em ibid.}
                {\bf B191}, 591(E) (1981).
\bibitem{Iof95} B.~L.~Ioffe, 
		Phys. Atom. Nucl.  {\bf 58}, 1408 (1995) [Yad. Fiz.  {\bf 58N8}, 1492 (1995)] [arXiv:hep-ph/		9501319].
\bibitem{Gas83} J.~Gasser and H.~Leutwyler,
                Ann. Phys. {\bf 158}, 142 (1984).
\bibitem{Col01} G.~Colangelo, J.~Gasser, and H.~Leutwyler,
                Nucl. Phys. {\bf B603}, 125 (2001) [arXiv:hep-ph/0103088].
\bibitem{Ovc88} A. A. Ovchinnikov and A. A. Pivovarov,
                Sov. J. Nucl. Phys. {\bf 48}, 721 (1988).
\bibitem{Glo95} L.~Y.~Glozman and D.~O.~Riska,
                Phys. Rep. {\bf 268}, 263 (1996)
                [arXiv:hep-ph/9505422].
\bibitem{Ris99} D.~O.~Riska and G.~E.~Brown,
                Nucl. Phys. {\bf A653}, 251 (1999)
                [arXiv:hep-ph/9902319].

\bibitem{Erkol}
G. Erkol, Th. A. Rijken, and R. G. E. Timmermans, in preparation.

\end{thebibliography}
\end{document}